\documentstyle{article}
\begin{document}
\begin{titlepage}
\large

\begin{center}
National Academy of Sciences of Ukraine\\
Institute for Condensed Matter Physics
\end{center}

\vspace{2cm}
\hspace{6cm}
\begin{tabular}{l}
Preprint \\
ICMP - 96 -  25E
\end{tabular}

\vspace{2cm}

\begin{center}
O.V. Derzhko, T.M. Verkholyak
\end{center}

\begin{center}
Spin-$\frac{1}{2}$ isotropic $XY$ chain
with Dzyaloshinskii-Moriya interaction in random lorentzian transverse field
\end{center}

\vspace{1.5cm}

\begin{abstract}
{The exact results for thermodynamical properties of one-di\-men\-sional
spin-$\frac{1}{2}$ isotropic $XY$ model with Dzya\-lo\-shin\-skii-Mo\-riya
interaction in random lorentzian transverse field are obtained. This
permits to discuss some approximate methods of disordered spin
systems theory. The approximate scheme of examining the thermodynamics
of one-dimensional spin-$\frac{1}{2}$ $XXZ$ Heisen\-berg mo\-del with
Dzya\-lo\-shin\-skii-Mo\-riya interaction in random loren\-tzi\-an field is
suggested.
}
\end{abstract}

\vspace{2cm}

\begin{center}
L'viv - 1996
\end{center}
\end{titlepage}

\noindent
Shortly after famous paper by E.Lieb, T.Schultz and D.Mattis \cite{lsm},
who making use of Jordan-Wigner transformation reformulated the Hamiltonian
of spin $s=\frac{1}{2}$ $XY$ chain in terms of non-interacting fermions and
obtained a series of rigorous results, the random versions of such models
attract much attention. Some exact results were derived by H.Nishimori
\cite{n} for isotropic $XY$ chain in random lorentzian transverse field. It
became possible because of the fact that after fermionization of such model
one comes to a system of electrons on lattice that may transfer from site
to site with the random (lorentzian) energy at sites. The average
one-particle
Green's function for such model was obtained first by P.Lloyd
\cite{l} (see also \cite{sai,ekl,zim,lgp}).

What follows is based on the notation that
similarly to \cite{n} one can consider the case of more complicate interspin
interaction including
to it the so-called Dzyaloshinskii-Moriya interaction. It was
introduced phenomenologically by I.E.Dzyaloshinskii \cite{d} and then derived
by T.Moriya \cite{m} and it is widely used as one of microscopical mechanism
(together with ANNNI model) of appearance of incommensurate phase in
crystals \cite{iz}. V.M.Kontorovich and V.M.Tsukernik noted
\cite{kc} that taking into account
of such interaction in $s=\frac{1}{2}$ $XY$ chain
does not destroy the
consideration proposed in Ref. \cite{lsm} since after Jordan-Wigner
transformation, as in the previous case, one comes to a quadratic in Fermi
operators form. In Ref. \cite{kc} the problem about the possibility of
appearance of spiral spin structure in such a chain was examined; for this
purpose the pair equal-time spin correlation functions were evaluated. Other
papers dealing with statistical mecanics of $s=\frac{1}{2}$ $XY$ chains with
Dzyaloshinskii-Moriya interaction \cite{zv1,zv2,dsm,dm1,dml,dlm,dm2,dm3}
like the Ref. \cite{kc} are devoted
to perfect (non-random) versions of the model.

The present paper contains
some exact results of statistical mechanics of $1D$ $s=\frac{1}{2}$
isotropic $XY$ model with Dzyaloshinskii-Moriya interaction in random
lorentzian field. The paper is organized as follows.
Fermionization and Lloyd's problem are considered in Section 1. Here the
average one-fermion Green's functions, the average spectral density and
the average
fermion correlation functions are derived. Section 2 contains calculations of
thermodynamical properties of the model in question. A discussion of the
estimation of static spin correlations is also given in it. The
comparison of exact results with the ones obtained within different
approximate approaches (Bose commutation rules approximation, Tyablikov-like
approximation, coherent potential approximation) are performed in Section 3.
The developed in Sections 1 and 2 scheme may be used for the approximate
study
of 1$D$ $s=\frac{1}{2}$ $XXZ$ Heisenberg model with Dzyaloshinskii-Moriya
interaction in random lorentzian field. This possibility is discussed in
Section 4. Conclusions are given in Section 5. Briefly the results of the
present paper were reported in \cite{der,dkv,dv}.\\

\renewcommand{\theequation}{\arabic{section}.\arabic{equation}}
\section{Fermionization, Lloyd's problem, average one-fermi\-on Gre\-en's
functions,average spectral density, average fer\-mi\-on correlation
functions}

A chain of $N$ spins $s=\frac{1}{2}$ with interaction between nearest
neighbours, that are in transverse fields with random component distributed
according to lorentzian law is considered. The Hamiltonian of the model
has the form
\begin{eqnarray}
&&
H=\sum_{j=1}^{N}(\Omega_0+\Omega_j)s_j^z+J\sum_{j=1}^{N-1}
(s^x_js^x_{j+1}+s^y_js^y_{j+1})
\nonumber\\
&&
+D\sum_{j=1}^{N-1}(s^x_js^y_{j+1}-s^y_js^x_{j+1})
\nonumber\\
&&
=\sum_{j=1}^{N}(\Omega_0+\Omega_j)\left(s_j^+s_j^--\frac{1}{2}\right)
\nonumber\\
&&
+\sum_{j=1}^{N-1}
\left(\frac{J+iD}{2}s^+_js^-_{j+1}+\frac{J-iD}{2}s^-_js^+_{j+1}\right).
\end{eqnarray}
After Jordan-Wigner transformation
\begin{eqnarray}
&&
c_1=s_1^-, c_j=(-2s_1^z)(-2s_2^z)...(-2s_{j-1}^z)s_j^-,\: j=2,...,N,
\nonumber\\
&&
c_1^+=s_1^+, c_j^+=(-2s_1^z)(-2s_2^z)...(-2s_{j-1}^z)s_j^+,\: j=2,...,N,
\nonumber\\
&&
\{c_i^+,c_j\}=\delta_{ij},\: \{c_i^+,c_j^+\}=0,\: \{c_i,c_j\}=0
\end{eqnarray}
one comes to the following quadratic in Fermi operators Hamiltonian
\begin{eqnarray}
&&
H=\sum_{j=1}^{N}(\Omega_0+\Omega_j)\left(c_j^+c_j-\frac{1}{2}\right)
\nonumber\\
&&
+\sum_{j=1}^{N-1}
\left(\frac{J+iD}{2}c^+_jc_{j+1}-\frac{J-iD}{2}c_jc^+_{j+1}\right),
\end{eqnarray}
that can be treated like in Ref. \cite{l}.

Let's introduce the following retarded and advanced temperature
two-times Green's functions \cite{zu}
\begin{eqnarray}
&&
G_{nm}^{\mp}(t)\equiv {\mp}i\theta({\pm}t)<\{c_n(t),c_m^+(0)\}>.
\end{eqnarray}
The goal of further consideration is to find the average Green's
functions
${\overline {G_{nm}^{\mp}(t)}}$,
where the average means
\begin{eqnarray}
&&
{\overline {(...)}}\equiv \int_{-\infty}^{+\infty}d\Omega_1...
\int_{-\infty}^{+\infty}d\Omega_N p(...,\Omega_j,...)(...),
\nonumber\\
&&
p(...,\Omega_j,...)=\prod_{j=1}^N\frac{1}{\pi}
\frac{\Gamma}{\Omega_j^2+\Gamma^2}
\end{eqnarray}
(i.e.
$\Omega_j$s
are independently distributed according to the lorentzian probability
distribution density centred at
$\Omega_j=0$
with the width $\Gamma$).

It is easy to get the
equation of motion for (1.4), namely
\begin{eqnarray}
&&
i\frac{d}{dt}G_{nm}^{\mp}(t)=\delta(t)\delta_{nm}+(\Omega_0+\Omega_n)
G_{nm}^{\mp}(t)
\nonumber\\
&&
+\frac{J+iD}{2}G_{n+1,m}^{\mp}(t)+
\frac{J-iD}{2}G_{n-1,m}^{\mp}(t).
\end{eqnarray}
Using spectral representation of Green's functions (1.4)
\begin{eqnarray}
&&
G_{nm}^{\mp}(t)=\frac{1}{2\pi}\int_{-\infty}^{+\infty}d\omega
\exp{(-i\omega t)}G_{nm}^{\mp}(\omega ),
\nonumber\\
&&
G_{nm}^{\mp}(\omega)=\int_{-\infty}^{+\infty}dt
\exp{(i\omega t)}G_{nm}^{\mp}(t)
\end{eqnarray}
the equations (1.6) can be rewritten in the form
\begin{eqnarray}
&&
\omega G_{nm}^{\mp}(\omega)=\delta_{nm}+(\Omega_0+\Omega_n)
G_{nm}^{\mp}(\omega)
\nonumber\\
&&
+\frac{J+iD}{2}G_{n+1,m}^{\mp}(\omega)+
\frac{J-iD}{2}G_{n-1,m}^{\mp}(\omega),
\end{eqnarray}
or in the form that is initial for locator expansion
\begin{eqnarray}
G_{nm}^{\mp}(\omega)=\frac{\delta_{nm}}{\omega-(\Omega_0+\Omega_n)}+
\frac{\frac{J+iD}{2}G_{n+1,m}^{\mp}(\omega)+
\frac{J-iD}{2}G_{n-1,m}^{\mp}(\omega)}{\omega-(\Omega_0+\Omega_n)}.
\end{eqnarray}
Formal solution as a series with respect to intersite interaction reads
\begin{eqnarray}
&&
G_{nm}^{\mp}(\omega)=\frac{\delta_{nm}}{\omega-(\Omega_0+\Omega_n)}
\nonumber\\
&&
+\frac{J+iD}{2}\frac{1}{\omega-(\Omega_0+\Omega_n)}
\frac{\delta_{n+1,m}}{\omega-(\Omega_0+\Omega_{n+1})}
\nonumber\\
&&
+\frac{J-iD}{2}\frac{1}{\omega-(\Omega_0+\Omega_{n})}
\frac{\delta_{n-1,m}}{\omega-(\Omega_0+\Omega_{n-1})}+...\;.
\end{eqnarray}

It is easy to average
$G_{nm}^{\mp}(\omega \pm i\epsilon)$
presented as (1.10). Really, one should
perform the averaging of
$[\frac{1}{\omega {\pm}i\epsilon-(\Omega_0+\Omega_j)}]^{k_j}$,
i.e. to calculate the
integral
\begin{eqnarray}
&&
\frac{1}{\pi}\int_{-\infty}^{+\infty}d\Omega_j\frac{\Gamma}{\Omega_j^2+
\Gamma^2}\left[\frac{1}{\omega {\pm}i\epsilon-
(\Omega_0+\Omega_j)}\right]^{k_j}
\nonumber\\
&&
=\frac{1}{\pi}\int_{-\infty}^{+\infty}d\Omega_j\frac{\Gamma}{(\Omega_j+
i\Gamma)(\Omega_j-i\Gamma)[\omega {\pm}i\epsilon-(\Omega_0+\Omega_j)]^{k_j}}.
\end{eqnarray}
For retarded (advanced) Green's function the integrand has one pole in the
lower (upper) half of the complex plane
$\Omega_j$
and two poles in the upper
(lower) half of the complex plane
$\Omega_j$. Thus expanding the contour of
integration in the lower (upper) half of the complex plane and using
the residuum theory
one ends up with
\begin{eqnarray}
&&
{\overline {\left[\frac{1}{\omega {\pm}i\epsilon-(\Omega_0+
\Omega_j)}\right]^{k_j}}}
=\left[\frac{1}{\omega {\pm}i\epsilon-(\Omega_0 \mp i\Gamma)}\right]^{k_j}
\end{eqnarray}
that is the famous property of lorentzian distribution. In result the average
series (1.10) can be summed up with the result
\begin{eqnarray}
&&
{\overline {G_{nm}^{\mp}(\omega \pm i\epsilon)}}=
\frac{\delta_{nm}}{\omega {\pm}i\epsilon-(\Omega_0 \mp i\Gamma)}
\nonumber\\
&&
+\frac{\frac{J+iD}{2}{\overline {G_{n+1,m}^{\mp}(\omega \pm i\epsilon)}}+
\frac{J-iD}{2}{\overline {G_{n-1,m}^{\mp}(\omega \pm i\epsilon)}}}
{\omega {\pm}i\epsilon-(\Omega_0 \mp i\Gamma)}.
\end{eqnarray}
Since the average Green's functions are translationally invariant the
equations (1.13) can be solved with the help of transformation
\begin{eqnarray}
&&
{\overline {G_{nm}^{\mp}(\omega \pm i\epsilon)}}=\frac{1}{N}\sum_{\kappa}
\exp[i(n-m){\kappa}]{\overline {G_{\kappa}^{\mp}(\omega \pm i\epsilon)}},
\end{eqnarray}
$\kappa =\frac{2\pi n}{N}, n=-\frac{N}{2}, -\frac{N}{2}+1,..., \frac{N}{2}-1$
for even $N$ or
$n=-\frac{N-1}{2}, -\frac{N-1}{2}+1,..., \frac{N-1}{2}$
for odd $N$;
the solution of algebraic equation for
${\overline {G_{\kappa}^{\mp}(\omega \pm i\epsilon)}}$
reads
\begin{eqnarray}
&&
{\overline {G_{\kappa}^{\mp}(\omega \pm i\epsilon)}}=\frac{1}
{\omega-[\Omega_0+\sqrt{J^2+D^2}\cos{(\kappa +\varphi)}]
{\pm}i(\epsilon+\Gamma)},
\nonumber\\
&&
\cos{\varphi}\equiv\frac{J}{\sqrt{J^2+D^2}},\;
\sin{\varphi}\equiv\frac{D}{\sqrt{J^2+D^2}}.
\end{eqnarray}
The average Green's functions in site representation can be found in result
of summation over
$\kappa$
in (1.14) that in thermodynamical limit reduces to the
following integration
\begin{eqnarray}
{\overline {G_{nm}^{\mp}(\omega \pm i\epsilon)}}=
\frac{1}{2\pi}\int_{-\pi}^{\pi}d\kappa \exp{[i(n-m)\kappa]}\times
\nonumber\\
\frac{1}{\omega -(\Omega_0+\frac{\sqrt{J^2+D^2}}{2}
\{\exp{[i(\kappa +\varphi)]}+\exp{[-i(\kappa +\varphi)]}\} \pm i
(\epsilon+\Gamma))}.
\end{eqnarray}
Setting
$z=\exp{[i(\kappa+\varphi)]}$
one comes to the contour integral on unit circle in complex plane $z$
\begin{eqnarray}
&&
{\overline {G_{nm}^{\mp}(\omega \pm i\epsilon)}}=
-\frac{\exp{[i\varphi (n-m)]}}{2\pi i}
\nonumber\\
&&
\times
\oint dz\frac{z^{n-m}}{\frac{\sqrt{J^2+D^2}}{2}z^2-
[\omega-\Omega_0{\pm}i(\epsilon+\Gamma)]z+\frac{\sqrt{J^2+D^2}}{2}}.
\end{eqnarray}
The denominator in the integrand in (1.17) should be presented as a product
$a(z-z_1)(z-z_2)$
where
\begin{eqnarray}
&&
a=\frac{\sqrt{J^2+D^2}}{2},
\nonumber\\
&&
z_{1}=\frac{\omega-\Omega_0{\pm}i(\epsilon+\Gamma)}{\sqrt{J^2+D^2}} +
\sqrt{\left[\frac{\omega-\Omega_0{\pm}i(\epsilon+\Gamma)}
{\sqrt{J^2+D^2}}\right]^2-1},
\nonumber\\
&&
z_{2}=\frac{\omega-\Omega_0{\pm}i(\epsilon+\Gamma)}{\sqrt{J^2+D^2}} -
\sqrt{\left[\frac{\omega-\Omega_0{\pm}i(\epsilon+\Gamma)}
{\sqrt{J^2+D^2}}\right]^2-1},
\end{eqnarray}
and since
$z_1z_2=1 \; \; \; z_1$($z_2$)
is over (in) the unit circle. For
$n\geq m$
the integration yields
\begin{eqnarray}
&&
\frac{\exp{[i\varphi (n-m)]}}{\sqrt{J^2+D^2}}
\frac{\{\frac{\omega-\Omega_0{\pm}i(\epsilon+\Gamma)}{\sqrt{J^2+D^2}}-
\sqrt{[\frac{\omega-\Omega_0{\pm}i(\epsilon+\Gamma)}{\sqrt{J^2+D^2}}]^2-1}\}
^{n-m}}
{\sqrt{[\frac{\omega-\Omega_0{\pm}i(\epsilon+\Gamma)}{\sqrt{J^2+D^2}}]^2-1}},
\end{eqnarray}
whereas for
$n\leq m$
\begin{eqnarray}
&&
\frac{\exp{[i\varphi (n-m)]}}{\sqrt{J^2+D^2}}
\frac{\{\frac{\omega-\Omega_0{\pm}i(\epsilon+\Gamma)}{\sqrt{J^2+D^2}}-
\sqrt{[\frac{\omega-\Omega_0{\pm}i(\epsilon+\Gamma)}{\sqrt{J^2+D^2}}]^2-1}\}
^{m-n}}
{\sqrt{[\frac{\omega-\Omega_0{\pm}i(\epsilon+\Gamma)}{\sqrt{J^2+D^2}}]^2-1}}.
\end{eqnarray}
Combing (1.19), (1.20) one
finally gets for (1.17)
\begin{eqnarray}
&&
{\overline {G_{nm}^{\mp}(\omega \pm i\epsilon)}}=
\nonumber\\
&&
\frac{\exp{[i\varphi (n-m)]}}{\sqrt{J^2+D^2}}
\frac{\{\frac{\omega-\Omega_0{\pm}i(\epsilon+\Gamma)}{\sqrt{J^2+D^2}}-
\sqrt{[\frac{\omega-\Omega_0{\pm}i(\epsilon+\Gamma)}{\sqrt{J^2+D^2}}]^2-1}\}
^{\mid n-m \mid }}
{\sqrt{[\frac{\omega-\Omega_0{\pm}i(\epsilon+\Gamma)}{\sqrt{J^2+D^2}}]^2-1}}.
\end{eqnarray}

The obtained result (1.21) gives the average elementary excitation
spectral density
\begin{eqnarray}
&&
{\overline {\rho(E)}}\equiv \frac{1}{N}
\sum_j{\overline {\delta(E-\Lambda_j)}}=
-\frac{1}{\pi}\frac{1}{N}\sum_\kappa
{\mbox {Im}}{\overline {G_{\kappa}^-(E+i\epsilon)}}
\nonumber\\
&&
=-\frac{1}{\pi} {\mbox {Im}} {\overline {G_{nn}^{-}(E)}}
=-\frac{1}{\pi} {\mbox {Im}}\frac{1}{\sqrt{(E-\Omega_0+i\Gamma)^2-(J^2+D^2)}}.
\end{eqnarray}
It will be used in the next Section for examination of thermodynamical
properties of the model in question.

In order to estimate spin correlations one should calculate average fermion
correlation function
${\overline {<c_m^+(0)c_n(t)>}}$
that can be found from the relation \cite{zu}
\begin{eqnarray}
&&
{\overline {<c_m^+(0)c_n(t)>}}=-\frac{1}{\pi}
{\mbox {Im}}\int_{-\infty}^{\infty} d\omega
\exp{(-i\omega t)}\frac{{\overline {G_{nm}^{-}(\omega+i\epsilon)}}}
{\exp{(\beta\omega)}+1}.
\end{eqnarray}
In particular, the static average fermion correlation function at low
tempereture limit comes out from the following calculation
\begin{eqnarray}
&&
{\overline {<c_{m}^+(0)c_{m+p}(0)>}}\equiv
{\overline {<c_{m}^+c_{m+p}>}}=
{\overline {<c_{n-p}^+(0)c_n(0)>}}
\nonumber\\
&&
=-\frac{1}{\pi}{\mbox {Im}}\int_{-\infty}^0 d\omega
{\overline {G_{n,n-p}^{-}(\omega+i\epsilon)}}
\nonumber\\
&&
=-\frac{1}{\pi}{\mbox {Im}}\exp{(i\varphi p)}\int_{-\infty}^{-\Omega}dy
\frac{[y+i\gamma-\sqrt{(y+i\gamma)^2-1}]^{\mid p\mid}}
{\sqrt{(y+i\gamma)^2-1}}
\nonumber\\
&&
=\frac{1}{\pi \mid p \mid}
{\mbox {Im}}
\left\{\exp{(i\varphi p)}\left[y+i\gamma-\sqrt{(y+i\gamma)^2-1}\right]
^{\mid p\mid}\mid_{y=-\infty}^{y=-\Omega}\right\}
\nonumber\\
&&
=\frac{1}{\pi \mid p \mid}
{\mbox {Im}}
\left\{\exp{(i\varphi p)}
\right.
\nonumber\\
&&
\left.
\times
\left[-\omega_0
+\sqrt{\frac{\sqrt
{(\omega_0^2-\gamma^2-1)^2+4\gamma^2\omega_0^2}+\omega_0^2-\gamma^2-1}{2}}
\right.\right.
\nonumber\\
&&
\left.\left.
+i\gamma-i\sqrt{\frac{\sqrt
{(\omega_0^2-\gamma^2-1)^2+4\gamma^2\omega_0^2}-
\omega_0^2+\gamma^2+1}{2}}\right]
^{\mid p\mid}\right\},
\end{eqnarray}
where
$y \equiv \frac{\omega}{\sqrt{J^2+D^2}}-\omega_0$,
$\omega_0 \equiv \frac{\Omega_0}
{\sqrt{J^2+D^2}}$,
$\gamma \equiv \frac{\epsilon+\Gamma}{\sqrt{J^2+D^2}}$.
The derived results (1.15), (1.21), (1.22), (1.24) remind the
corresponding expressions obtained in slightly different cases in Refs.
\cite{zv3,zv4}.

\setcounter{equation}{0}
\section{
Thermodynamical properties. The influence of Dzya\-lo\-shin\-skii-Moriya
interaction}

The obtained in the previous Section results are of great use in
understanding the
thermodynamical properties of the model in question. Really, consider a model
with a certain realization of transverse fields at sites. After exploiting
Jordan-Wigner transformation (1.2) for the Hamiltonian (1.1) one comes to a
quadratic in Fermi operators form that can be diagonalized by a canonical
transformation
$\eta_k=\sum^N_{j=1}(g_{kj}c_j+h_{kj}c_j^+)$
\cite{lsm} (see also \cite{dk1,dk2}) with the
result
$H=\sum_{k=1}^N \Lambda_k(\eta_k^+\eta_k-\frac{1}{2}).$
Elementary excitation spectrum
$\Lambda_k$
and the coefficients $g_{kj}$ and $h_{kj}$ are determined from
$\Lambda_k g_{kn}=\sum_{i=1}^N g_{ki}A_{in}$,
$-\Lambda_k h_{kn}=\sum_{i=1}^N h_{ki}A_{in}^*$,
where
$A_{ij}=(\Omega_0+\Omega_i)\delta_{ij}+
\frac{J+iD}{2}\delta_{j,i+1}+\frac{J-iD}{2}\delta_{j,i-1}$.
The calculation of free energy per site for this realization is
straightforward
\begin{eqnarray}
&&
f=\lim_{N \rightarrow \infty }\frac{1}{N}\left\{-\frac{1}{\beta} \ln{}
\prod_k\left[\exp{\left(-\frac{\beta \Lambda_k}{2}\right)}+
\exp{\left(\frac{\beta \Lambda_k}{2}\right)}
\right]\right\}
\nonumber\\
&&
=-\frac{1}{\beta} \int dE \rho (E)
\ln{\left(2\cosh{\frac{\beta E}{2}}\right)},
\nonumber\\
&&
\rho (E)\equiv \lim_{N \rightarrow \infty }\frac{1}{N}
\sum_k\delta(E-\Lambda_k)
\end{eqnarray}
and the result of its averaging over configuration is given by
\begin{eqnarray}
&&
{\overline f}=-\frac{1}{\beta} \int dE {\overline {\rho (E)}}
\ln{\left(2\cosh{\frac{\beta E}{2}}\right)}
\nonumber\\
&&
=-\frac{1}{\beta} \int dE {\overline {\rho (E)}} \ln{[1+\exp{(-\beta E)}]}-
\frac{\Omega_0}{2},
\end{eqnarray}
where
${\overline {\rho (E)}}$
is the average spectral density that has been found in the
Section 1 (formula (1.22)).

Further one finds the internal energy
\begin{eqnarray}
&&
{\overline e}={\overline f}+
\beta \frac{\partial {\overline f}}{\partial \beta}=
\int dE {\overline {\rho (E)}} \frac{E}{1+\exp{(\beta E)}}-
\frac{\Omega_0}{2},
\end{eqnarray}
the entropy
\begin{eqnarray}
&&
{\overline s}=\beta^2\frac{\partial {\overline f}}{\partial \beta}=
\int dE {\overline {\rho (E)}}
\left\{\ln{[1+\exp{(-\beta E)}]}+\frac{\beta E}{1+\exp{(\beta E)}}\right\},
\end{eqnarray}
and the specific heat
\begin{eqnarray}
&&
{\overline c}=-\beta \frac{\partial {\overline s}}{\partial \beta}=
\beta^2 \int dE {\overline {\rho (E)}}
\frac{E^2}{(2\cosh{\frac{\beta E}{2}})^2}.
\end{eqnarray}
Using magical property of
${\overline {\rho (E)}} $
(1.22)
\begin{eqnarray}
&&
\frac{\partial}{\partial \Omega_0}{\overline {\rho (E)}}=
\frac{\partial}{\partial \Omega_0}\left[-\frac{1}{\pi}{\mbox {Im}}\frac{1}
{\sqrt{(E-\Omega_0+i\Gamma)^2-(J^2+D^2)}}\right]
\nonumber\\
&&
=
-\frac{\partial}{\partial E}{\overline {\rho (E)}}
\end{eqnarray}
one finds the transverse magnetization
\begin{eqnarray}
&&
{\overline {<\frac{1}{N} \sum_{j=1}^Ns_j^z>}}=
\frac{\partial {\overline f}}{\partial \Omega_0}=
\int dE {\overline {\rho (E)}} \frac{1}{1+\exp{(\beta E)}}-\frac{1}{2}
\end{eqnarray}
and static transverse susceptibility
\begin{eqnarray}
&&
{\overline {\chi_{zz}}}=
\frac{\partial {\overline {<\frac{1}{N} \sum_{j=1}^Ns_j^z>}}}
{\partial \Omega_0}=-\beta \int dE {\overline {\rho (E)}}
\frac{1}{(2\cosh{\frac{\beta E}{2}})^2}.
\end{eqnarray}
Note, that since at $T\rightarrow 0$
${\frac{1}{1+\exp{(\beta E)}}}{\rightarrow 0}$ if $E>0$ and
${\frac{1}{1+\exp{(\beta E)}}}{\rightarrow 1}$ if $E<0$,
the transverse magnetization
and static transverse susceptibility at $\Omega_0=0$
in low tempereture limit are given by
\begin{eqnarray}
&&
{\overline {<\frac{1}{N} \sum_{j=1}^Ns_j^z>}}
\nonumber\\
&&
=\int_0^{-\Omega_0} dE' \left[-\frac{1}{\pi} {\mbox {Im}} \frac{1}
{\sqrt{(E'+i\Gamma)^2-(J^2+D^2)}}\right],
\end{eqnarray}
\begin{eqnarray}
&&
{\overline {\chi_{zz}}}=-\frac{1}{\pi}
\frac{1}{\sqrt{\Gamma^2+J^2+D^2}}.
\end{eqnarray}

The results of numerical investigation of thermodynamical properties
of the model (1.1) are depicted in Figs.1-5. The dependence on
Dzyaloshinskii-Moriya interaction
$(D=0,~ D=0.5J,~ D=J) $
of the average spectral density
${\overline {\rho (E)}}\;(\Omega_0=0) $
for different values of the width of lorentzian distribution
$\Gamma\; (\Gamma =0,\;\Gamma  =0.25J,\;\Gamma =0.5J,\;\Gamma =J)$
is shown in Fig.1. In Figs.2,3 it is shown the temperature behaviour of
entropy and specific heat for
$\Omega_0=0$
for different values of Dzyaloshinskii-Moriya interaction
$(D=0,~ D=0.5J,~ D=J) $
and different values of
$\Gamma\; (\Gamma =0,\;\Gamma =J).$
The changes in the
dependence of transverse magnetization on the value of transverse field
$\Omega_0$
that are caused by Dzyaloshinskii-Moriya interaction
$(D=0,~ D=0.5J,~ D=J) $
for several values of
$\Gamma\; (\Gamma =0,\;\Gamma =J)$ at $T=0$
can be seen in Fig.4. In Fig.5 the temperature behaviour of static
transverse susceptibility in zero transverse
field for few values of Dzyaloshinskii-Moriya interaction
$(D=0,~ D=0.5J,~ D=J) $
and
$\Gamma\; (\Gamma =0,\;\Gamma =J)$
are presented.

As it can be easily seen from the formula for
${\overline {\rho (E)}}$
(1.22) and in Fig.1 the presence of Dzyaloshinskii-Moriya interaction
\begin{figure}[tbhp]
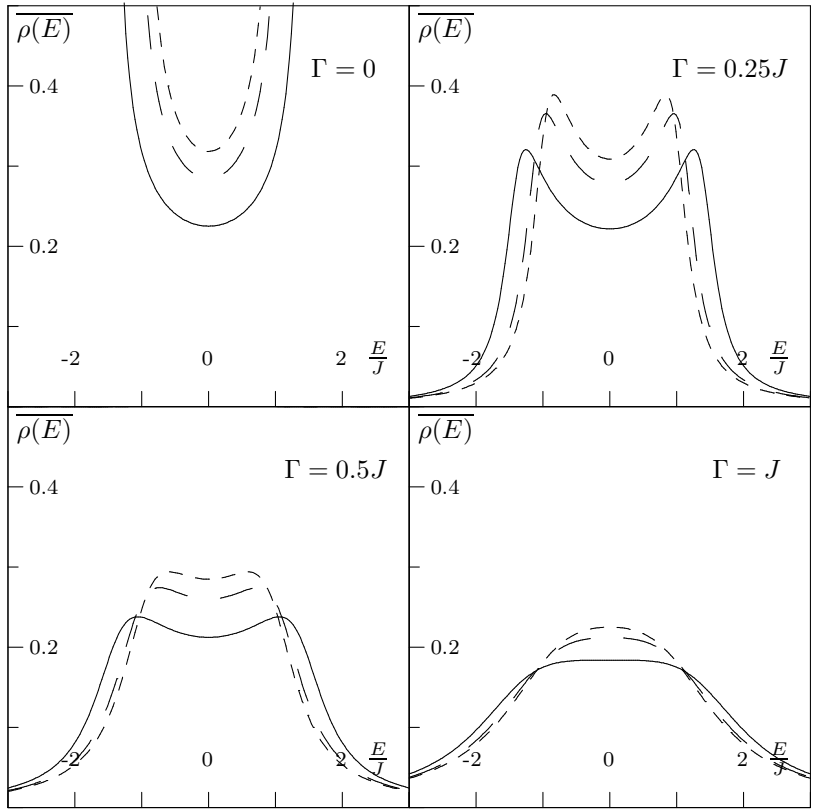

\input F25E1.PIC
\caption{The average spectral density ${\overline {\rho (E)}}$ (1.22) vs.
$\frac{E}{J}$ for
different values of $\Gamma$; $\Omega_0=0$, $D=0$ (dashed curves),
$D=0.5J$ (long dashed curves) and $D=J$ (solid curves).}
\end{figure}
formally causes the change of
$J^2$
to
$J^2+D^2$.
This leads to effective increase of interspin interaction and results in
broadening of zones both with sharp edges (when
$\Gamma=0$) and with smooth ones
because of randomness (when
$\Gamma \neq 0$).
Rather small quantitative changes in temperature behaviour of entropy and
specific heat (Figs.2,3) reveal the tendency caused by
\begin{figure}[hp]
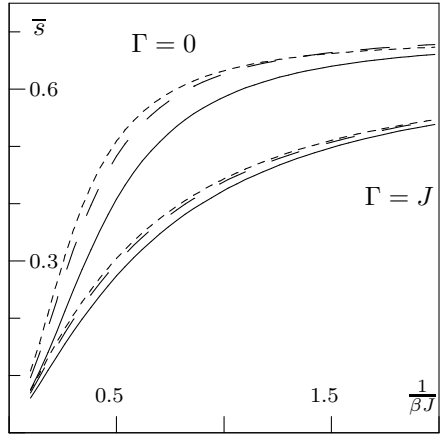

\begin{center}
\input F25E2.PIC
\end{center}
\caption{Temperature dependence of entropy ${\overline s}$ for $\Gamma=0$
and $\Gamma=J$; $\Omega_0=0$, $D=0$ (dashed curves), $D=0.5J$ (long
dashed curves) and $D=J$ (solid curves).} \end{figure}
\begin{figure}[hp]
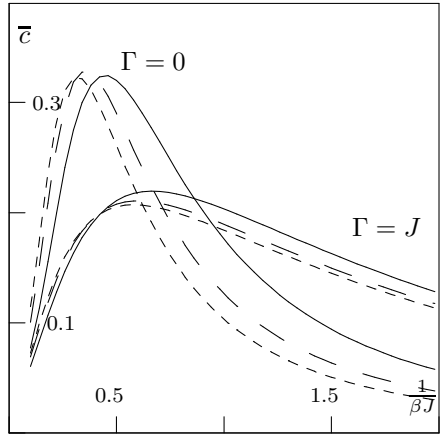

\begin{center}
\input F25E3.PIC
\end{center}
\caption{Temperature dependence of specific heat ${\overline c}$
for $\Gamma=0$ and $\Gamma=J$; $\Omega_0=0$, $D=0$ (dashed curves),
$D=0.5J$ (long dashed curves) and $D=J$ (solid curves).}
\end{figure}
\begin{figure}[hp]
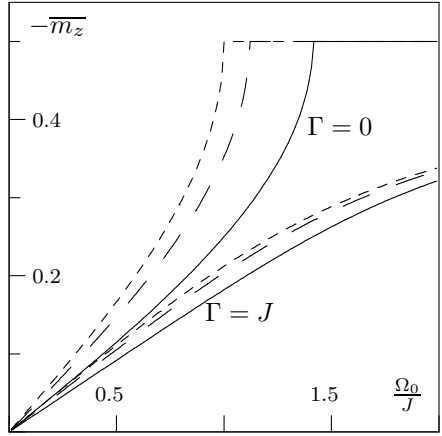

\begin{center}
\input F25E4.PIC
\end{center}
\caption{The dependence of transverse magnetization
${\overline {m_z}}\equiv
{\overline {<\frac{1}{N} \sum_{j=1}^Ns_j^z>}}$ on transverse field
$\frac{\Omega_0}{J}$
for $\Gamma=0$ and $\Gamma=J$; $1/\beta=0$, $D=0$ (dashed curves),
$D=0.5J$ (long dashed curves) and $D=J$ (solid curves).}
\end{figure}
\begin{figure}[hp]
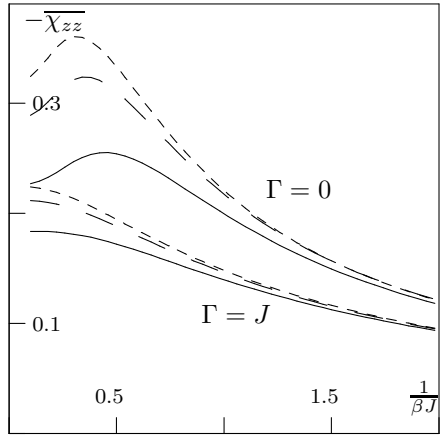

\begin{center}
\input F25E5.PIC
\end{center}
\caption{Temperature dependence of static transverse susceptibility
${\overline {\chi_{zz}}}$ for $\Gamma=0$
and $\Gamma=J$; $\Omega_0=0$, $D=0$ (dashed curves),
$D=0.5J$ (long dashed curves) and $D=J$ (solid curves).}
\end{figure}
Dzyaloshinskii-Moriya interaction. Dzyaloshinskii-Moriya interaction
decreases the transverse magnetization for a given value of transverse
field in both cases when $\Gamma=0$
and when
$\Gamma \neq 0$
(Fig.4) and decreases the value of static transverse susceptibility in zero
transverse field in low temperature region. The obtained result are in
agreement with the derived ones for non-random version of 1$D$
$s=\frac{1}{2}$ isotropic $XY$ model in transverse field \cite{dm2,dm3}.

Unfortunately, the obtained results do not permit to get any exact
estimations for static spin correlation functions. Considering the
simplest one
${\overline {<s_j^z s_{j+n}^z>}}$,
using the relation
$s_j^z=c_j^+ c_j -\frac{1}{2}$
and exploiting Wick-Bloch-de Dominicis theorem one finds
\begin{eqnarray}
&&
{\overline {<s_j^z s_{j+n}^z>}}
\nonumber\\
&&
={\overline {<c_j^+ c_jc_{j+n}^+c_{j+n}>}}
-
\frac{1}{2}{\overline {<c_j^+ c_j>}}-
\frac{1}{2}{\overline {<c_{j+n}^+ c_{j+n}>}}+\frac{1}{4}
\nonumber\\
&&
={\overline {<c_j^+ c_j><c_{j+n}^+c_{j+n}>}}-
{\overline {<c_j^+ c_{j+n}^+><c_jc_{j+n}>}}
\nonumber\\
&&
+{\overline {<c_j^+ c_{j+n}><c_jc_{j+n}^+>}}-
{\overline {<c_j^+ c_j>}}+\frac{1}{4}.
\end{eqnarray}
Thus, knowing only
${\overline {<c_m^+ c_n>}}$ it is not possible to derive any rigorous
results even for simplest (transverse) static spin correlation function.
It is worthwhile to note that its calculation faces with the similar problems
that were considered in \cite{sai}. Apparently, the problem of spin
correlations can be solved within developed recently numerical approach
\cite{dk1,dk2,dk3,dk4,dk5}.
\setcounter{equation}{0}
\section{
Standard approximate approaches in the theory of disordered spin systems}

Let's consider the results one faces with after adopting some standard
approximate approaches in disordered spin systems theory while considering
the thermodynamics of the model defined by (1.1).

\subsection{
Bose commutation rules approximation for spin operators $s^+,\:s^-$}

Implying instead of Pauli commutation rules for spin operators
$s^+$, $s^-$ $[s_j^-,s_m^+]=\delta_{jm}(1-2s_j^+s_j^-)$
Bose commutation rules, that is
\begin{eqnarray}
&&
[s_j^-,s_m^+]\approx \delta_{jm},
\end{eqnarray}
one comes to a system of bosons on lattice that may transfer from site to
site with random (lorentzian) energy at sites.
Introducing the following Green's functions
\begin{eqnarray}
&&
D_{nm}^{\mp}(t)\equiv {\mp}i\theta({\pm}t)<[s_n^-(t),s_m^+(0)]>,
\end{eqnarray}
and repeating the derivation of Section 1 one ends up with
\begin{eqnarray}
&&
{\overline {D_{nm}^{\mp}(\omega{\pm}i\epsilon)}}=
\nonumber\\
&&
\frac{\exp{[i\varphi (n-m)]}}{\sqrt{J^2+D^2}}
\frac{\{\frac{\omega-\Omega_0{\pm}i(\epsilon+\Gamma)}{\sqrt{J^2+D^2}}-
\sqrt{[\frac{\omega-\Omega_0{\pm}i(\epsilon+\Gamma)}{\sqrt{J^2+D^2}}]^2-1}\}
^{\mid n-m \mid }}
{\sqrt{[\frac{\omega-\Omega_0{\pm}i(\epsilon+\Gamma)}{\sqrt{J^2+D^2}}]^2-1}}
\end{eqnarray}
that gives for average spectral density formula (1.22). However, the
average boson correlation function
${\overline {<s_m^+(0)s_n^-(t)>}}=-\frac{1}{\pi}
{\mbox {Im}}\int_{-\infty}^{\infty} d\omega\times$ $
\exp{(-i\omega t)}\frac{{\overline {D_{nm}^{-}(\omega+i\epsilon)}}}
{\exp{(\beta\omega)}-1}$
contains the denominator that tends to 0 when
$\omega\rightarrow 0.$

Let's discuss this problem in details. After assuming Bose commutation
relation one has the quadratic in Bose operators
$s_j^+,\;s_j^-$
form (1.1) that with the help of linear canonical transformation
$\gamma_k=\sum_{j=1}^N(f_{kj}s_j^-+d_{kj}s_j^+)$
can be
diagonalized with the result
$H=-\frac{1}{2}\sum_{j=1}^N(\Omega_0+\Omega_j)+
\sum_{k=1}^N {\cal E}_k\gamma_k^+\gamma_k+const$.
Since ${\cal E}_k$, $f_{kj}$, $d_{kj}$
are determined from the equations
${\cal E}_k f_{kn}=\sum_{i=1}^N f_{ki}A_{in}$,
$-{\cal E}_k d_{kn}=\sum_{i=1}^N d_{ki}A_{in}^*$,
one immediately concludes that
${\cal E}_k=\Lambda_k $.
The ground state energy of this Hamiltonian coincides with the exact value if
$const=-\frac{1}{2}\sum_{k=1}^N\Lambda_k+\frac{1}{2}\sum_{j=1}^N(\Omega_0
+\Omega_j)$
and thus finally
$H=\sum_{k=1}^N \Lambda_k(\gamma_k^+\gamma_k-\frac{1}{2})$.
Free energy per particle is given by
\begin{eqnarray}
&&
f=\lim_{N \rightarrow \infty}\frac{1}{N}\times
\nonumber\\
&&
\left\{-\frac{1}{\beta} \ln{}
\prod_{k}
\exp{\left(\frac{\beta \Lambda_{k}}{2}\right)}
[1+\exp{(-\beta \Lambda_{k})}+\exp{(-2\beta \Lambda_{k})}
+...]\right\}
\nonumber\\
&&
\begin{array}{c}
{\mbox {all }} \Lambda_k>0\\
=\\
\\
\end{array}
\lim_{N \rightarrow \infty}\frac{1}{N}\left[-\frac{1}{\beta} \sum_{k}
\ln{}\frac{\exp{\left(\frac{\beta \Lambda_{k}}{2}\right)}}
{1-\exp{(-\beta \Lambda_{k})}}\right]
\nonumber\\
&&
=\frac{1}{\beta}\int dE \rho (E) \ln{[1-\exp{(-\beta E)}]}-
\frac{1}{2}\int dE \rho (E)E.
\end{eqnarray}
Note, that the partition function for Bose system exists only when
the condition all
$\Lambda_{\kappa}>0$
(or
$\rho (E)=0$
if
$E\leq 0$)
is valid. Really, otherwise
the state without bosons is not the ground state, since one-boson state
has smaller energy, the energy of
two-bosons state is more smaller etc., the probability of their
appearance increases respectively and thus the
partition function tends to infinity. This difficulty
arises because of approximate treating of elementary excitations as
bosons (they are exactly fermionic objects in the case
under consideration) and it is crushing if
$\rho (E)\neq 0$
for
$E\leq 0.$
Thus, one can consider the average thermodynamical quantities obtained within
approximation (3.1), that is, free energy
\begin{eqnarray}
&&
{\overline f}=\frac{1}{\beta}
\int dE {\overline {\rho (E)}} \ln{[1-\exp{(-\beta E)}]}-\frac{\Omega_0}{2},
\end{eqnarray}
internal energy
\begin{eqnarray}
&&
{\overline e}=
\int dE {\overline {\rho (E)}} \frac{E}{\exp{(\beta E)}-1}
-\frac{\Omega_0}{2},
\end{eqnarray}
entropy
\begin{eqnarray}
&&
{\overline s}=
\int dE {\overline {\rho (E)}}
\left\{-\ln{[1-\exp{(-\beta E)}]}+\frac{\beta E}{\exp{(\beta E)}-1}\right\},
\end{eqnarray}
specific heat
\begin{eqnarray}
&&
{\overline c}=
\beta^2 \int dE {\overline {\rho (E)}}
\frac{E^2}{(2\sinh{\frac{\beta E}{2}})^2},
\end{eqnarray}
transverse magnetization
\begin{eqnarray}
&&
{\overline {<\frac{1}{N} \sum_{j=1}^Ns_j^z>}}=
\int dE {\overline {\rho (E)}} \frac{1}{\exp{(\beta E)}-1}-\frac{1}{2},
\end{eqnarray}
and static transverse susceptibility
\begin{eqnarray}
&&
{\overline {\chi_{zz}}}=
-\beta \int dE {\overline {\rho (E)}}
\frac{1}{(2\sinh{\frac{\beta E}{2}})^2}
\end{eqnarray}
when the relation
\begin{eqnarray}
&&
{\overline {\rho (E)}}=0\; {\mbox {for}} \; E\leq 0
\end{eqnarray}
holds true.

Considering at first the case
$\Gamma =0 $,
when
\begin{eqnarray}
\rho (E){=}\left\{\begin{array}{cl}
\!\!\!\frac{1}{\pi}\frac{1}{\sqrt{J^2+D^2-(E-\Omega_0)^2}}, &\!\!\!
\Omega_0{-}\sqrt{J^2{+}D^2}{<}E{<}\Omega_0{+}
\sqrt{J^2{+}D^2},\!\!\!\!\!\!\!\!\!\!\!\!
\\
0, & \mbox{otherwise},
\end{array}\right.
\end{eqnarray}
one finds that (3.11) is valid only in the case of strong transverse fields
\begin{eqnarray}
&&
\Omega_0 > \sqrt{J^2+D^2};
\end{eqnarray}
Dzyaloshinskii-Moriya interaction increases the value of the field behind
which (3.11) becomes true. In the case $\Gamma \neq 0 $
one immediately finds that (3.11) is never true, since there always will be
elementary excitations with negative energy and
therefore it is impossible to treat them as bosons.
Thus, the consideration of the disordered version of the model
demands the revision of the problem of validity of approximation that was
suitable for study of non-random version of the model.

The results of numerical calculation of temperature behaviours of entropy and
specific heat according to exact formulae (2.4), (2.5) (solid lines)
and approximate ones (3.7),
(3.8) (long dashed curves) in the case of validity of approximation (3.1)
$(\Gamma =0,~\Omega_0 =(\sqrt{2}+0.1)J,~D=0,~D=0.5J,~D=J)$
are presented in Figs.6,7. These results show that Bose commutation rules
approximation for spin operators
$s^+,\; s^-$
gives suitable results only for low temperatures, and in the presence of
Dzyaloshinskii-Moriya interaction only at very low temperatures.

\subsection{
 Tyablikov-like approximation}

Does not assuming Bose commutation rules for operators
$s^+,\; s^-$
(3.1), one faces with the equations of motion that contain more complicated
Green's functions
\begin{eqnarray}
&&
i\frac{d}{dt}D_{nm}^{\mp}(t)=\delta(t)\delta_{nm}(-2<s_n^z>)
+(\Omega_0+\Omega_n)D_{nm}^{\mp}(t)
\nonumber\\
&&
+\frac{J+iD}{2}(-2)({\mp}i\theta ({\pm}t)<[s_n^z(t)s_{n+1}^-(t),s_m^+]>)
\nonumber\\
&&
+\frac{J-iD}{2}(-2)({\mp}i\theta ({\pm}t)<[s_{n-1}^-(t)s_n^z(t),s_m^+]>).
\end{eqnarray}
\begin{figure}[hp]
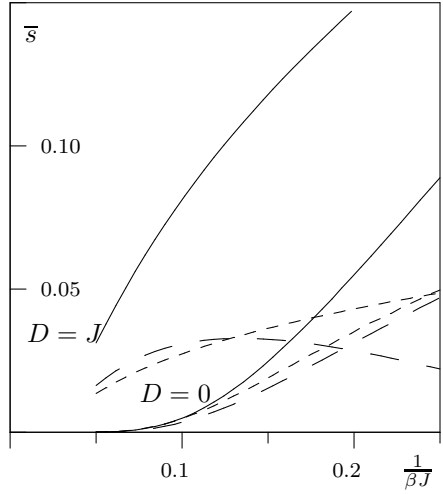

\begin{center}
\input F25E6.PIC
\end{center}

\vspace{-5mm}
\caption{Entropy vs. temperature for $\Gamma=0$,
$\Omega_0 =(\protect\sqrt{2}+0.1)J$:
exact results (solid curves), Bose commutation rules approximation (long
dashed curves), Tyablikov-like approximation (dashed curves).}
\end{figure}

\begin{figure}[hp]
\begin{center}
\input F25E7.PIC
\end{center}

\vspace{-5mm}
\caption{Specific heat vs. temperature for $\Gamma=0$,
$\Omega_0 =(\protect\sqrt{2}+0.1)J$:
exact results (solid curves), Bose commutation rules approximation (long
dashed curves), Tyablikov-like approximation (dashed curves).}
\end{figure}

Within Tyablikov-like approximation it is supposed that
\begin{eqnarray}
&&
<s_n^z(t)s_{n+1}^-(t)s_m^+> \approx {\overline {<s^z>}}<s_{n+1}^-(t)s_m^+>,
\nonumber\\
&&
<s_m^+s_n^z(t)s_{n+1}^-(t)> \approx {\overline {<s^z>}}<s_m^+s_{n+1}^-(t)>,
\nonumber\\
&&
<s_{n-1}^-(t)s_n^z(t)s_m^+> \approx {\overline {<s^z>}}<s_{n-1}^-(t)s_m^+>,
\nonumber\\
&&
<s_m^+s_{n-1}^-(t)s_n^z(t)> \approx {\overline {<s^z>}}<s_m^+s_{n-1}^-(t)>.
\end{eqnarray}
Then instead of (3.14) one has
\begin{eqnarray}
i\frac{d}{dt}D_{nm}^{\mp}(t)=-2{\overline {<s^z>}}\delta(t)\delta_{nm}
+(\Omega_0+\Omega_n)D_{nm}^{\mp}(t)
\nonumber\\
+(-2{\overline {<s^z>}})\frac{J+iD}{2}D_{n+1,m}^{\mp}(t)
+(-2{\overline {<s^z>}})\frac{J-iD}{2}D_{n-1,m}^{\mp}(t).
\end{eqnarray}
Acting like in Section 1 one ends up with
\begin{eqnarray}
{\overline {D_{nm}^{\mp}(\omega{\pm}i\epsilon)}}=
\nonumber\\
\frac{\exp{[i\varphi (n{-}m)]}}{\sqrt{J^2{+}D^2}}
\frac{\{\frac{\omega-\Omega_0{\pm}i(\epsilon+\Gamma)}
{(-2{\overline {<s^z>}})\sqrt{J^2+D^2}}{-}
\sqrt{[\frac{\omega-\Omega_0{\pm}i(\epsilon+\Gamma)}
{(-2{\overline {<s^z>}})\sqrt{J^2+D^2}}]^2{-}1}\}
^{\mid n-m \mid }}
{\sqrt{[\frac{\omega-\Omega_0{\pm}i(\epsilon+\Gamma)}
{(-2{\overline {<s^z>}})\sqrt{J^2+D^2}}]^2-1}}
\end{eqnarray}
that yields the following result for average spectral density
\begin{eqnarray}
&&
{\overline {\rho(E)}}
=-\frac{1}{\pi} {\mbox {Im}}\frac{1}{\sqrt{[\frac{E-\Omega_0+i\Gamma}
{(-2{\overline {<s^z>}})}]^2-(J^2+D^2)}}.
\end{eqnarray}
The introduced average transverse magnetization at site is determined from
the equation
\begin{eqnarray}
&&
{\overline {<s^z>}}={\overline {<s^+s^->}}-\frac{1}{2}=
-\frac{1}{\pi} {\mbox {Im}}\int_{-\infty}^{\infty}d\omega
\frac{{\overline {D_{mm}^-(\omega+i\epsilon)}}}{\exp{(\beta\omega)}-1}-
\frac{1}{2}
\nonumber\\
&&
=\int_{-\infty}^{\infty}\frac{d\omega}{\exp{(\beta\omega)}-1}
{\overline {\rho (\omega)}}-\frac{1}{2}.
\end{eqnarray}

Equation (3.19) contains Bose factor
$\frac{1}{\exp{(\beta\omega)}-1}$
and thus, apparently, Tyab\-likov-like approximation is possible if
${\overline {\rho (\omega)}}=0$ for $\omega \leq 0$.
The temperature behaviours of entropy and specific heat found within
Tyablikov-like
approximation (3.15) (dashed lines) in comparison with exact results
(solid lines) and results obtained
within approximation (3.1) (long dashed lines) are shown in Figs.6,7.
Despite some improvment over the Bose commutation rules approximation
that can be seen in Figs.6,7
the results obtained within Tyablikov-like
approximation generally speaking are not closer to exact ones in comparison
with the results derived within
Bose commutation rules approximation.

\subsection{
Randomly disordered crystals theory methods
}

Let's consider 1$D$ spin-$\frac{1}{2}$ isotropic $XY$ model with
Dzya\-lo\-shin\-skii-Mo\-riya interaction in random (not necessary
lorentzian)
transverse field within usually used methods in the theory of disordered
crystals \cite{ekl,zim,hk}. The starting point is the equation for Green's
functions that after introducing the notations
\begin{eqnarray}
&&
W_{pr}\equiv\Omega_p\delta_{pr},~
V_{ps}\equiv\frac{J+iD}{2}\delta_{s,p+1}+\frac{J-iD}{2}\delta_{s,p-1}
\end{eqnarray}
can be written in the form
\begin{eqnarray}
&&
\omega G_{nm}^{\mp}(\omega)=\delta_{nm}+\Omega_0 G_{nm}^{\mp}(\omega)+
W_{nr}G_{rm}^{\mp}(\omega)+V_{ns}G_{sm}^{\mp}(\omega)
\end{eqnarray}
(for $G_{nm}^{\mp}(\omega)$ defined by (1.4) or (3.2) within Bose commutation
rules approximation) or
\begin{eqnarray}
&&
\omega D_{nm}^{\mp}(\omega)=-2<s_n^z>\delta_{nm}+
\Omega_0 D_{nm}^{\mp}(\omega)+
W_{nr}D_{rm}^{\mp}(\omega)
\nonumber\\
&&
+(-2<s_n^z>)V_{ns}D_{sm}^{\mp}(\omega)
\end{eqnarray}
(for $D_{nm}^{\mp}(\omega)$ (3.2) within Tyablikov approximation, $<s_n^z>$
is only thermodynamically averaged value (without configurational
averaging as in (3.15)) of transverse spin at site $n$);
the summation over the repeating indices from $1$ to $N$ is implied.
The different approaches of randomly disordered crystals theory are
constructed from so called propagator and locator expansions.

{\em Propagator expansion. }
Let's rewrite the Hamiltonian of the system in question (1.1)
in the form
\begin{eqnarray}
&&
H=\ ^aH+\sum_{j=1}^N\Omega_js_j^z=\ ^aH+\sum_{j=1}^N\Omega_j\left(s_j^+s_j^-
-\frac{1}{2}\right)
\end{eqnarray}
and introduce Green's functions $^a G_{nm}^{\mp}(\omega )$ or
$^a D_{nm}^{\mp}(\omega )$ for the system with Hamiltonian
$^a H$ that, naturally satisfy the following equations
\begin{eqnarray}
&&
\omega \ ^aG_{nm}^{\mp}(\omega)=\delta_{nm}+\Omega_0 \ ^aG_{nm}^{\mp}(\omega)
+V_{ns} \ ^aG_{sm}^{\mp}(\omega)
\end{eqnarray}
or
\begin{eqnarray}
&&
\omega \ ^aD_{nm}^{\mp}(\omega)=-2<s_n^z>\delta_{nm}+
\Omega_0 \ ^aD_{nm}^{\mp}(\omega)
\nonumber\\
&&
+(-2<s_n^z>)V_{ns} \ ^aD_{sm}^{\mp}(\omega).
\end{eqnarray}
Multiplying (3.21) (or (3.22)) by $\ ^a G_{gn}^{\mp}(\omega )$
($\ ^a D_{gn}^{\mp}(\omega )$) one comes to the following equations
\begin{eqnarray}
&&
G_{gm}^{\mp}(\omega )=\ ^aG_{gm}^{\mp}(\omega)+^aG_{gn}^{\mp}(\omega)
W_{nr} G_{rm}^{\mp}(\omega)
\end{eqnarray}
or
\begin{eqnarray}
&&
D_{gm}^{\mp}(\omega )=\ ^aD_{gm}^{\mp}(\omega)\frac{<s_n^z>}{<s^z>}
+\ ^aD_{gn}^{\mp}(\omega)
\tilde{W}_{nr} D_{rm}^{\mp}(\omega),
\nonumber\\
&&
\tilde{W}_{nr}\equiv \frac{W_{nr}}{-2<s^z>}+V_{nr}
\left(\frac{<s_n^z>}{<s^z>}-1\right).
\end{eqnarray}
Note, that since generally speaking $<s_n^z> \neq <s^z>$ even in the
case of diagonal disorder after
Tyablikov-like approximation one faces with non-diagonal disorder problem.
Expanding of
(3.26) (or (3.27)) in degrees of $W_{nr}(\tilde{W}_{nr})$ leads to propagator
expansion.

{\em Locator expansion. }
Let's introduce locators
\begin{eqnarray}
&&
g_{nm}^{\mp}(\omega )\equiv g_n^{\mp}\delta_{nm}\equiv
\frac{1}{\omega-(\Omega_0+\Omega_n)}\delta_{nm}
\end{eqnarray}
or
\begin{eqnarray}
&&
d_{nm}^{\mp}(\omega )\equiv d_n^{\mp}\delta_{nm}\equiv
\frac{-2<s^z_n>}{\omega-(\Omega_0+\Omega_n)}\delta_{nm}.
\end{eqnarray}
Then equations (3.21), (3.22) can be rewritten in the form
\begin{eqnarray}
&&
G_{nm}^{\mp}(\omega)=g_{nm}^{\mp}(\omega)+g_{np}^{\mp}(\omega)V_{ps}
G_{sm}^{\mp}(\omega)
\end{eqnarray}
or
\begin{eqnarray}
&&
D_{nm}^{\mp}(\omega)=d_{nm}^{\mp}(\omega)+d_{np}^{\mp}(\omega)V_{ps}
D_{sm}^{\mp}(\omega).
\end{eqnarray}
While expanding r.h.s. of (3.30), (3.31) in degrees of
$V_{ps}$ one comes to locator
expansion.

Further analysis deals only with diagonal disorder when, for example,
the propagator expansions have the form
\begin{eqnarray}
G_{gm}^{\mp}(\omega)=
\ ^aG_{gm}^{\mp}(\omega)+
\nonumber\\
\ ^aG_{gn}^{\mp}(\omega)
\Omega_n\ ^aG_{nm}^{\mp}(\omega)+ \ ^aG_{gn}^{\mp}(\omega)\Omega_n
\ ^aG_{np}^{\mp}(\omega)\Omega_p \ ^aG_{pm}^{\mp}(\omega)+...
\end{eqnarray}
or
\begin{eqnarray}
D_{gm}^{\mp}(\omega)=
\ ^aD_{gm}^{\mp}(\omega)+
\nonumber\\
\ ^aD_{gn}^{\mp}(\omega)
\tilde{\Omega}_n \ ^aD_{nm}^{\mp}(\omega)+\ ^aD_{gn}^{\mp}(\omega)
\tilde{\Omega}_n
\ ^aD_{np}^{\mp}(\omega)\tilde{\Omega}_p \ ^aD_{pm}^{\mp}(\omega)+...\;,
\nonumber\\
\tilde{\Omega}_n=-\Omega_n /2{\overline {<s^z>}}.
\end{eqnarray}
Extracting in (3.32) $t$-matrix
\begin{eqnarray}
&&
t_n\equiv \frac{\Omega_n}{1-\ ^aG_{nn}^{\mp}(\omega)\Omega_n},
\end{eqnarray}
one can rewrite (3.32) as a series in degrees of $t$-matrix
\begin{eqnarray}
G_{gm}^{\mp}(\omega)= \ ^aG_{gm}^{\mp}(\omega)+
\ ^aG_{gn}^{\mp}(\omega)
\Omega_n \ ^aG_{nm}^{\mp}(\omega)
\nonumber\\
+ \ ^aG_{gn}^{\mp}(\omega)\Omega_n
\ ^aG_{nn}^{\mp}(\omega)\Omega_n \ ^aG_{nm}^{\mp}(\omega)
\nonumber\\
+\ ^aG_{gn}^{\mp}(\omega)\Omega_n
\ ^aG_{np~(n\neq p)} ^{\mp}(\omega)\Omega_p \ ^aG_{pm}^{\mp}(\omega)
\nonumber\\
+\ ^aG_{gn}^{\mp}(\omega)\Omega_n
\ ^aG_{nn}^{\mp}(\omega)\Omega_n \ ^aG_{nn}^{\mp}(\omega)\Omega_n
\ ^aG_{nm}^{\mp}(\omega)
\nonumber\\
+\ ^aG_{gn}^{\mp}(\omega)\Omega_n
\ ^aG_{np~(n\neq p)} ^{\mp}(\omega)\Omega_p \ ^aG_{pp}^{\mp}(\omega)\Omega_p
\ ^aG_{pm}^{\mp}(\omega)
\nonumber\\
+\ ^aG_{gn}^{\mp}(\omega)\Omega_n
\ ^aG_{nn}^{\mp}(\omega)\Omega_n \ ^aG_{nf~(n\neq f)} ^{\mp}(\omega)\Omega_f
\ ^aG_{fm}^{\mp}(\omega)
\nonumber\\
+\ ^aG_{gn}^{\mp}(\omega)\Omega_n
\ ^aG_{np~(n\neq p)} ^{\mp}(\omega)\Omega_p
\ ^aG_{pf~(p\neq f)} ^{\mp}(\omega)\Omega_f \ ^aG_{fm}^{\mp}(\omega)+...
\nonumber\\
=\ ^aG_{gm}^{\mp}(\omega)+
\ ^aG_{gn}^{\mp}(\omega)t_n \ ^aG_{nm}^{\mp}(\omega)
\nonumber\\
+\ ^aG_{gn}^{\mp}(\omega)t_n \ ^aG_{np~(n\neq p)} ^{\mp}(\omega)t_p
\ ^aG_{pm}^{\mp}(\omega)+...\;.
\end{eqnarray}
Within approximation of average $t$-matrix one assumes that
\begin{eqnarray}
&&
t_n\simeq {\overline t} \equiv \int d\Omega_1...d\Omega_N
p(\Omega_1,...,\Omega_N) \frac{\Omega_n}{1-\ ^aG_{nn}^{\mp}(\omega)\Omega_n},
\end{eqnarray}
where
\begin{eqnarray}
&&
{\ ^aG_{nm}^{\mp}(\omega)}=
\nonumber\\
&&
\frac{\exp{[i\varphi (n-m)]}}{\sqrt{J^2+D^2}}
\frac{[\frac{\omega-\Omega_0{\pm}i\epsilon}{\sqrt{J^2+D^2}}-
\sqrt{(\frac{\omega-\Omega_0{\pm}i\epsilon}{\sqrt{J^2+D^2}})^2-1}]
^{\mid n-m \mid }}
{\sqrt{(\frac{\omega-\Omega_0{\pm}i\epsilon}{\sqrt{J^2+D^2}})^2-1}}.
\end{eqnarray}
In result one is able to sum the series for average Green's functions (3.35)
\begin{eqnarray}
&&
{\overline {G_{gm}^{\mp}(\omega)}}=
\left(\frac{^aG^{\mp}(\omega)}{1-\ ^aG^{\mp}(\omega){\overline t}}
\right)_{gm}.
\end{eqnarray}

Within coherent potential approximation one should seek the Green's functions
of the system  with Hamiltonian
\begin{eqnarray}
\check{H}=
\nonumber\\
\sum_{j=1}^{N}(\Omega_0+\check{\Omega})s_j^z{+}J
\sum_{j=1}^{N-1}(s^x_js^x_{j+1}{+}
s^y_js^y_{j+1}){+}D\sum_{j=1}^{N-1}(s^x_js^y_{j+1}{-}s^y_js^x_{j+1}),
\end{eqnarray}
where
$\check{\Omega}$
is unknown coherent field. These Green's functions are given by
\begin{eqnarray}
&&
{\check{G}_{nm}^{\mp}(\omega)}=
\nonumber\\
&&
\frac{\exp{[i\varphi (n-m)]}}{\sqrt{J^2+D^2}}
\frac{[\frac{\omega-\Omega_0-\check{\Omega}{\pm}i\epsilon}{\sqrt{J^2+D^2}}-
\sqrt{(\frac{\omega-\Omega_0-\check{\Omega}{\pm}i\epsilon}
{\sqrt{J^2+D^2}})^2-1}]
^{\mid n-m \mid }}
{\sqrt{(\frac{\omega-\Omega_0-\check{\Omega}{\pm}i\epsilon}
{\sqrt{J^2+D^2}})^2-1}}.
\end{eqnarray}
Since
\begin{eqnarray}
&&
H=\check{H}+\sum_{j=1}^N(\Omega_j-\check{\Omega})\left(s_j^+s_j^-
-\frac{1}{2}\right),\\
&&
\omega \check{G}_{nm}^{\mp}(\omega)=
\delta_{nm}+(\Omega_0+\check{\Omega})\check{G}_{nm}^{\mp}(\omega)+
V_{ns}\check{G}_{sm}^{\mp}(\omega),
\end{eqnarray}
one can get acting like while deriving (3.26) the following equation
\begin{eqnarray}
&&
G_{gm}^{\mp}(\omega)=\check{G}_{gm}^{\mp}(\omega)+
\check{G}_{gn}^{\mp}(\omega)\check{W}_{nr}G_{rm}^{\mp}(\omega),
\nonumber\\
&&
\check{W}_{nr}=W_{nr}-\check{\Omega}\delta_{nr}=
(\Omega_n-\check{\Omega})\delta_{nr},
\end{eqnarray}
and hence the following propagator expansion
\begin{eqnarray}
&&
G_{gm}^{\mp}(\omega)=\check{G}_{gm}^{\mp}(\omega)+
\check{G}_{gn}^{\mp}(\omega)
(\Omega_n-\check{\Omega})\check{G}_{nm}^{\mp}(\omega)
\nonumber\\
&&
+\check{G}_{gn}^{\mp}(\omega)(\Omega_n-\check{\Omega})
\check{G}_{np}^{\mp}(\omega)
(\Omega_p-\check{\Omega})\check{G}_{pm}^{\mp}(\omega)+...\;.
\end{eqnarray}
This series can be rewritten as an expansion in degrees of
$\check{t}$
-matrix
\begin{eqnarray}
&&
\check{t}_n\equiv \frac{\Omega_n-\check{\Omega}}
{1-\check{G}_{nn}^{\mp}(\omega)(\Omega_n-\check{\Omega})};
\end{eqnarray}
namely,
\begin{eqnarray}
G_{gm}^{\mp}(\omega)=
\nonumber\\
\check{G}_{gm}^{\mp}(\omega){+}\check{G}_{gn}^{\mp}(\omega)
\check{t}_n\check{G}_{nm}^{\mp}(\omega){+}
\check{G}_{gn}^{\mp}(\omega)\check{t}_n
\check{G}_{np~(n \neq p)}^{\mp}(\omega)
\check{t}_p\check{G}_{pm}^{\mp}(\omega){+}...\;.
\end{eqnarray}
Determining the coherent field
$\check{\Omega}$
from the condition
\begin{eqnarray}
&&
{\overline {\check{t}_n}}\equiv \int d\Omega_1...d\Omega_N
p(\Omega_1,...,\Omega_N)
\frac{\Omega_n-\check{\Omega}}
{1-\check{G}_{nn}^{\mp}(\omega)(\Omega_n-\check{\Omega})}=0,
\end{eqnarray}
where accorging to (3.40)
\begin{eqnarray}
&&
\check{G}_{nn}^{\mp}(\omega\pm i\epsilon)
=\frac{1}{\sqrt{(\omega-\Omega_0-\check{\Omega}
{\pm}i{\epsilon})^2-(J^2+D^2)}},
\end{eqnarray}
one finds that
\begin{eqnarray}
&&
{\overline {G_{gm}^{\mp}(\omega)}}=
\check{G}_{gm}^{\mp}(\omega)+\check{G}_{gn}^{\mp}(\omega)
{\overline {\check{t}_n}}\check{G}_{nm}^{\mp}(\omega)
\nonumber\\
&&
+\check{G}_{gn}^{\mp}(\omega)
{\overline {\check{t}_n\check{G}_{np~(n \neq p)}^{\mp}(\omega)\check{t}_p}}
\check{G}_{pm}^{\mp}(\omega)+...\simeq
\check{G}_{gm}^{\mp}(\omega),
\end{eqnarray}
that is the desired result within coherent potential approximation.

In the case of lorentzian transverse field the equation for coherent field
$\check{\Omega}$
(3.47) reads
\begin{eqnarray}
&&
\int_{-\infty}^{\infty}d\Omega_j\frac{1}{\pi}\frac{\Gamma}{(\Omega_j+i\Gamma)
(\Omega_j-i\Gamma)}
\nonumber\\
&&
\times \frac{(\Omega_j-\check{\Omega})
\sqrt{(\omega-\Omega_0-\check{\Omega}{\pm}i{\epsilon})^2-(J^2+D^2)}}
{\sqrt{(\omega-\Omega_0-\check{\Omega}{\pm}i{\epsilon})^2-(J^2+D^2)}-
\Omega_j+\check{\Omega}}=0.
\end{eqnarray}
Supposing that
${\mbox {Im}}[\check{\Omega}+\sqrt{(\omega-\Omega_0-\check{\Omega}
{\pm}i{\epsilon})^2-
(J^2+D^2)}]
\begin{array}{c}
> \\
<
\end{array}
0$
one can perform integration with the help of the residuum theory getting
in result instead of (3.50)
\begin{eqnarray}
&&
\frac{({\mp}i\Gamma-\check{\Omega})
\sqrt{(\omega-\Omega_0-\check{\Omega}{\pm}i{\epsilon})^2-(J^2+D^2)}}
{\sqrt{(\omega-\Omega_0-\check{\Omega}{\pm}i{\epsilon})^2-(J^2+D^2)}
+\check{\Omega}{\pm}i\Gamma}=0.
\end{eqnarray}
Equation (3.51) has solutions
$\check{\Omega}={\mp}i\Gamma$
and after insertion them into (3.40) one gets exact result (1.21). It can
be proved
{\em post priory} the possibility of assumed displacement of poles in (3.50)
at least for
$\omega\rightarrow \infty .$
Thus, in the case of lorentzian transverse field the coherent potetial
approximation contains exact result for average Green's functions
${\overline {G_{nm}^{\mp}(\omega)}}.$

Consider now another version of random transverse field that is given by
probability density
\begin{eqnarray}
&&
p(\Omega_1,...,\Omega_N)=\prod_{j=1}^N[x\delta(\Omega_j)+
(1-x)\delta(\Omega_j-{\mho})].
\end{eqnarray}
Then the equation for coherent field
$\check{\Omega}$
(3.47) reads
\begin{eqnarray}
&&
x\frac{-\check{\Omega}}{1-\check{G}_{nn}^{\mp}(\omega)(-\check{\Omega})}+
(1-x)\frac{{\mho}-\check{\Omega}}{1-\check{G}_{nn}^{\mp}(\omega)
({\mho}-\check{\Omega})}=0
\end{eqnarray}
and after some calculation reduces to 3th order algebraic equation for
$\check{\Omega}$
\begin{eqnarray}
\check{\Omega}^3+
\nonumber\\
\frac{(x^2{-}2x)\mho^2{-}(J^2{+}D^2){+}4(1{-}x)\mho
(\omega{-}\Omega_0){+}(\omega{-}\Omega_0)^2}{2(x\mho{+}\Omega_0{-}\omega)}
\check{\Omega}^2+
\nonumber\\
\frac{(1{-}x)\mho(J^2{+}D^2){-}(1{-}x)^2\mho^2(\omega{-}\Omega_0)
{-}(1{-}x)\mho(\omega {-}\Omega_0)^2}{x\mho{+}
\Omega_0{-}\omega}\check{\Omega}+
\nonumber\\
\frac{-(1-x)^2\mho^2(J^2+D^2)+(1-x)^2\mho^2(\omega -
\Omega_0)^2}{2(x\mho+\Omega_0-\omega)}
\nonumber\\
=\check{\Omega}^3+a\check{\Omega}^2+b\check{\Omega}+c=0.
\end{eqnarray}
It is generally-known \cite{kk} that at first one should substitute
$\check{\Omega}=y-\frac{a}{3}$
obtaining in result
$y^3+py+q=0$
with
$p=-\frac{a^2}{3}+b,\:q=2\left(\frac{a}{3}\right)^3-\frac{ab}{3}+c.$
Then the real "noncomlete" cubic equation has
\begin{itemize}
\item {one real and two conjugate complex roots if
$Q>0$;}
\item{
three real roots at least two of which coincide if
$Q=0;$}
\item{
three different real roots if
$Q<0;$}
\end{itemize}
here
$Q\equiv \left(\frac{p}{3}\right)^3+ \left(\frac{q}{2}\right)^2.$
The solution can be presented in the trigonometrical form
\begin{itemize}
\item if
$Q\geq 0,\:p>0,$
then
\begin{eqnarray}
y_1=-2\sqrt{\frac{p}{3}}\cot{2\alpha},\;
y_{2,3}=\sqrt{\frac{p}{3}}\left(\cot{2\alpha}{\pm}i\sqrt{3}\csc{2\alpha}
\right),
\nonumber\\
\tan{\alpha}=\sqrt[3]{\tan{\frac{\beta}{2}}}\: \left({\mid}\alpha{\mid}
\leq\frac{\pi}{4}\right),
\: \tan{\beta}=\frac{2}{q}\sqrt{\left(\frac{p}{3}\right)^3}\:
\left({\mid}\beta{\mid}\leq\frac{\pi}{2}\right);
\end{eqnarray}
\item if
$Q\geq 0,\:p<0,$
then
\begin{eqnarray}
y_1=-2\sqrt{-\frac{p}{3}}\csc{2\alpha},\;
y_{2,3}=\sqrt{-\frac{p}{3}}\left(\csc{2\alpha}{\pm}i\sqrt{3}
\cot{2\alpha}\right),
\nonumber\\
\tan{\alpha}=\sqrt[3]{\tan{\frac{\beta}{2}}}\:
\left({\mid}\alpha{\mid}{\leq}\frac{\pi}{4}\right),
\: \sin{\beta}=\frac{2}{q}\sqrt{\left(-\frac{p}{3}\right)^3}\:
\left({\mid}\beta{\mid}{\leq}\frac{\pi}{2}\right);
\end{eqnarray}
\item if
$Q<0$
(and hence
$p<0$
), then
\begin{eqnarray}
y_1=2\sqrt{-\frac{p}{3}}\cos{\frac{\alpha}{3}},\;
y_{2,3}=-2\sqrt{-\frac{p}{3}}\cos{\left(\frac{\alpha}{3}{\pm}\frac{\pi}{3}
\right)},
\nonumber\\
\cos{\alpha}=-\frac{q}{2\sqrt{\left(-\frac{p}{3}\right)^3}}
\end{eqnarray}
\end{itemize}
(all cubic roots are real).

Further one should act in a following way:
1) for a given $x$ to calculate $a,\: b,\: c$ in (3.54), $p,\: q$ and $Q$;
2) comparing $Q$ and $p$ with the zero, to write down
$y_{1,2,3};$
and hence
$\check{\Omega}_{1,2,3}$;
3) to check whether
${\check {\overline t}}$
after inserting
$\check{\Omega}_j$
is really equal to zero (since the algebraic transformation from (3.53)
to (3.54) may lead to appearance of extra roots);
4) to substitute
$\check{\Omega_j}$
into $\check{G}_{nn}^-(E+i\epsilon)$ (3.40);
and 5) to calculate the average spectral density within coherent
potential approximation (3.47)
\begin{eqnarray}
{\overline {\rho(E)}}\simeq -\frac{1}{\pi} {\mbox {Im}}\check{G}_{nn}^-(E)=
-\frac{1}{\pi}{\mbox {Im}}\frac{1}{\sqrt{(E-\Omega_0{-}\check{\Omega}_j
{+}i\epsilon)^2{-}
(J^2{+}D^2)}}.
\end{eqnarray}

Further the case $J=1,\:D=0,\:\mho=1$ is under consideration.
Here the example of performing of this program is given:
\vspace{1cm}\\
\vspace{0.5cm}
$x=0.01,\\
E=0.02,\\
a=47.015001,\; b=-97.000202,\; c=48.985399,\\
p=-833.803646,\; q=9267.099096,\; Q=0.896875,\\
\check {\Omega}_1=-49.014405  +i0.000000,\\
\check {\Omega}_2=0.999702  +i0.001967,\\
\check {\Omega}_3=0.999702  -i0.001967,\\
{\mid} \overline {\check t}{\mid}_{\check {\Omega}_1}
{=}1.012523{\times} 10^{-13},\;
{\mid} \overline {\check t}{\mid}_{\check {\Omega}_2}
{=}3.978199{\times} 10^{-3},\;
\vspace{0.5cm}
{\mid} \overline {\check t}{\mid}_{\check {\Omega}_3}
{=}3.064551{\times} 10^{-12};\\
E=0.03,\\
a=22.505001,\; b=-47.985301,\; c=24.480448,\\
p=-216.810318,\; q=1228.762879,\; Q=0.087101,\\
\check {\Omega}_1=-24.504032  +i0.000000,\\
\check {\Omega}_2=0.999516  +i0.002358,\\
\check {\Omega}_3=0.999516  -i0.002358,\\
{\mid} \overline {\check t}{\mid}_{\check {\Omega}_1}
{=}1.187939{\times} 10^{-14},\;
{\mid} \overline {\check t}{\mid}_{\check {\Omega}_2}
{=}4.811497{\times} 10^{-3},\;
\vspace{0.5cm}
{\mid} \overline {\check t}{\mid}_{\check {\Omega}_3}
{=}2.295226{\times} 10^{-12};\\
E=0.04,\\
a=14.331667,\; b=-31.640401,\; c=16.308864,\\
p=-100.105961,\; q=385.512229,\; Q=0.023724,\\
\check {\Omega}_1=-16.330344  +i0.000000,\\
\check {\Omega}_2=0.999339  +i0.002665,\\
\check {\Omega}_3=0.999339  -i0.002665,\\
{\mid} \overline {\check t}{\mid}_{\check {\Omega}_1}
{=}5.717649{\times} 10^{-15},\;
{\mid} \overline {\check t}{\mid}_{\check {\Omega}_2}
{=}5.488112{\times} 10^{-3},\;
\vspace{0.5cm}
{\mid} \overline {\check t}{\mid}_{\check {\Omega}_3}
{=}2.070775{\times} 10^{-12};\\
E=0.05,\\
a=10.242501,\; b=-23.463002,\; c=12.220623,\\
p=-58.432610,\; q=171.921947,\; Q=0.009691,\\
\check {\Omega}_1=-12.240840  +i0.000000,\\
\check {\Omega}_2=0.999170  +i0.002918,\\
\check {\Omega}_3=0.999170  -i0.002918,\\
{\mid} \overline {\check t}{\mid}_{\check {\Omega}_1}
{=}1.362799{\times} 10^{-14},\;
{\mid} \overline {\check t}{\mid}_{\check {\Omega}_2}
{=}6.062763{\times} 10^{-3},\;
\vspace{0.5cm}
{\mid} \overline {\check t}{\mid}_{\check {\Omega}_3}
{=}3.098188{\times} 10^{-12},\\$
etc. Since
$\check {\overline t}$ with
$\check {\Omega}_1$
for $x=0$ and $x=1$ is not equal to zero, this root has been rejected.
Unfortunately, there are no exact results for a random model in question
(3.52).
Nevertheless, for arbitrary random spin-$\frac{1}{2}$ anisotropic $XY$ model
in transverse field
it is possible to calculate numerically (see \cite{dk1,dk2}) the quantity
$R(E^2)=\frac{1}{N}\sum_{k=1}^{N}\delta(E^2-\Lambda_k^2),$
the average value
of which is connected with the average value of
$\rho (E)$ by the relation
\begin{eqnarray}
&&
{\overline {R(E^2)}}=\frac{{\overline {\rho (E)}}+{\overline {\rho (-E)}}}
{2\mid E\mid}.
\end{eqnarray}

In Fig.8 the results of calculations within coherent potential
\begin{figure}[hp]
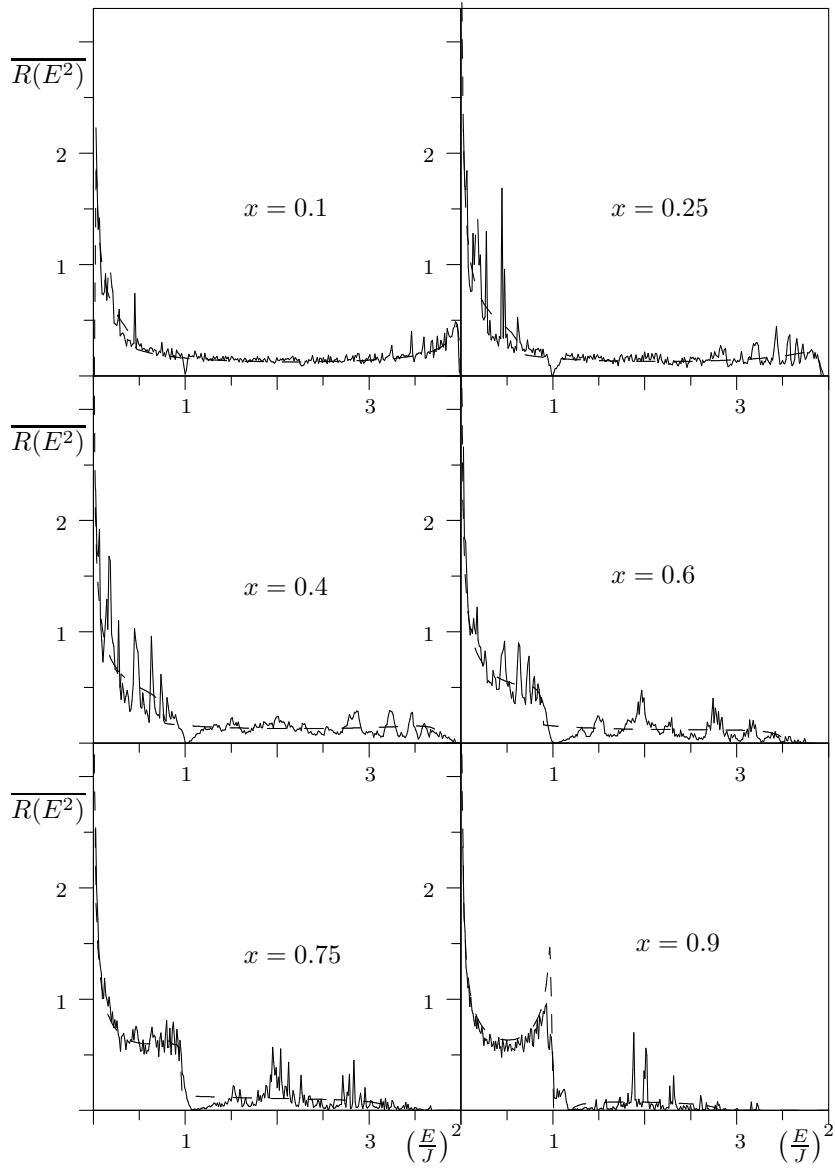

\input F25E8.PIC
\caption{$\overline{R(E^2)}$ vs. $E^2$: exact results (solid curves)
(the averaging is done only over few random realizations)
and the results within coherent potential approximation
(long dashed curves) for the model
with disorder (3.52):
$\Omega_0=0$, $J=1$, $D=0$, $\mho =1$.}
\end{figure}
approxi\-mation (3.58), (3.59) (broken lines) are depicted together with the
exact results (solid lines).
The comparison shows just how little is the change in $\overline {R(E^2)}$
and thus in thermodynamical quantities.
This seems to be conditioned
by the fact that for the model in question
(since it is described by Hamiltonian (1.3))
the thermodynamical averaging has been
performed exactly.

\setcounter{equation}{0}
\section{
One-dimensional spin-$\frac{1}{2}$ $XXZ$ Heisenberg model with
Dzyaloshinskii-Moriya interaction in lorent\-zian random external field}

This Section is devoted to examining of somewhat more complicated than (1.1)
case. Namely, now the intersite interaction contains the coupling of $z$
components of neighbouring spins and the Hamiltonian reads
\begin{eqnarray}
H{=}\sum_{j=1}^{N}(\Omega_0{+}\Omega_j)s_j^z{+}J
\sum_{j=1}^{N-1}(s^x_js^x_{j+1}{+}
s^y_js^y_{j+1}){+}D\sum_{j=1}^{N-1}(s^x_js^y_{j+1}{-}s^y_js^x_{j+1})
\nonumber\\
+J^z\sum_{j=1}^{N-1}s_j^zs_{j+1}^z
=\sum_{j=1}^{N}(\Omega_0+\Omega_j)\left (s_j^+s_j^--\frac{1}{2}\right )
\nonumber\\
+\sum_{j=1}^{N-1}
\left (\frac{J+iD}{2}s^+_js^-_{j+1}+\frac{J-iD}{2}s^-_js^+_{j+1}\right )
\nonumber\\
+J^z\sum_{j=1}^{N-1}\left(s_j^+s_j^-s_{j+1}^+s_{j+1}^-
-\frac{1}{2}s_{j}^+s_{j}^-
-\frac{1}{2}s_{j+1}^+s_{j+1}^-+\frac{1}{4}\right).
\end{eqnarray}
After Jordan-Wigner transformation (1.2) one gets
\begin{eqnarray}
&&
H=-\frac{N\Omega_0}{2}+\frac{(N-1)J^z}{4}-\frac{1}{2}\sum_{j=1}^{N}\Omega_j+
\nonumber\\
&&
\sum_{j=1}^{N}
\left [\Omega_0+\Omega_j-\frac{J^z}{2}(2-\delta_{j,1}-\delta_{j,N})\right ]
c_j^+c_j+
\nonumber\\
&&
\sum_{j=1}^{N-1}
\left (\frac{J+iD}{2}c^+_jc_{j+1}{-}\frac{J-iD}{2}c_jc^+_{j+1}\right )
{+}J^z\sum_{j=1}^{N-1}c_j^+c_jc_{j+1}^+c_{j+1}.
\end{eqnarray}


\noindent
Note, that  the key difficulty is that due to new intersite interaction one
faces with the terms that are the products of four Fermi operators.
It is easy to find the equations of motion for Green's functions (1.4)
\begin{eqnarray}
&&
i\frac{d}{dt} G_{nm}^{\mp}(t)=
\delta (t)\delta_{nm}+(\Omega_0+\Omega_n-J^z)
G_{nm}^{\mp}(t)
\nonumber\\
&&
+\frac{J+iD}{2}G_{n+1,m}^{\mp}(t)+\frac{J-iD}{2}G_{n-1,m}^{\mp}(t)
\nonumber\\
&&
+J^z [{\mp}i\theta ({\pm}t)<\{(c_nc_{n+1}^+c_{n+1})(t),c_m^+(0)\}>
\nonumber\\
&&
{\mp}i\theta ({\pm}t)<\{(c_{n-1}^+c_{n-1}c_n)(t),c_m^+(0)\}>].
\end{eqnarray}

Equations (4.3) contain higher Green's functions and thus cannot be solved
exactly. Nevertheless, making the approximation
\begin{eqnarray}
&&
<\{(c_{n+1}^+c_{n+1}c_n)(t),c_m^+(0)\}>\approx {\overline {<c_j^+c_j>}}
<\{c_n(t),c_m^+\}>
\nonumber\\
&&
-{\overline {<c_{j+1}^+c_j>}} <\{c_{n+1}(t),c_m^+\}>,
\nonumber\\
&&
<\{(c_{n-1}^+c_{n-1}c_n)(t),c_m^+(0)\}>\approx {\overline {<c_j^+c_j>}}
<\{c_n(t),c_m^+\}>
\nonumber\\
&&
-{\overline {<c_{j-1}^+c_j>}} <\{c_{n-1}(t),c_m^+\}>,
\end{eqnarray}
one gets instead of (4.3) the equations
\begin{eqnarray}
i\frac{d}{dt} G_{nm}^{\mp}(t)=\delta (t)\delta_{nm}+[\Omega_0
+J^z(2{\overline {<c_j^+c_j>}}-1)+\Omega_n]G_{nm}^{\mp}(t)+
\nonumber\\
\frac{J{-}2J^z{\overline {<c_{j+1}^+c_j>}}{+}iD}{2}G_{n+1,m}^{\mp}(t)
{+}\frac{J{-}2J^z{\overline {<c_{j-1}^+c_j>}}{-}iD}{2}G_{n-1,m}^{\mp}(t)
\end{eqnarray}
that can be treated like (1.6). Substituting
$\Omega_0+J^z(2{\overline {<c_j^+c_j>}}-1)$
instead of
$\Omega_0,\:J-2J^z{\overline {<c_{j\pm 1}^+c_j>}}\pm iD$
instead of
$J\pm iD$
into (1.15) one finds that
\begin{eqnarray}
&&
{\overline {G_{\kappa}^{\mp}(\omega{\pm}i\epsilon)}}=
\frac{1}{\omega-\left [{\overline {\overline {\Omega_0}}}+J_+\exp{(i\kappa)}
+J_-\exp{(-i\kappa)}\right ] {\pm}i(\epsilon+\Gamma)},
\nonumber\\
&&
{\overline {\overline {\Omega_0}}}\equiv \Omega_0+
J^z(2{\overline {<c_j^+c_j>}}-1),
\nonumber\\
&&
J_{\pm}\equiv
\frac{J-2J^z{\overline {<c_{j{\pm}1}^+c_j>}}{\pm}iD}{2}.
\end{eqnarray}

Returning back to the site representation needs the calculation of
the following integral
(compare with (1.16)-(1.21))
\begin{eqnarray}
&&
{\overline {G_{nm}^{\mp}(\omega{\pm}i\epsilon)}}=\frac{1}{2\pi}
\int_{-\pi}^{\pi}d\kappa \exp{[i(n-m)\kappa]}
\nonumber\\
&&
\times
\frac{1}{\omega-\left [{\overline {\overline {\Omega_0}}}+J_+\exp{(i\kappa)}
+J_-\exp{(-i\kappa)}\right ] {\pm}i(\epsilon+\Gamma)}.
\end{eqnarray}
Putting
$z=\exp{(\pm i\kappa)}$
(for
$n\geq m$
and
$n\leq m,$
respectively) one comes to contour integral over unit circle centred at
$z=0$
in complex plane $z$
\begin{eqnarray}
&&
{\overline {G_{nm}^{\mp}(\omega{\pm}i\epsilon)}}=-\frac{1}{2\pi i}
\oint dz\frac{z^{n-m}}
{J_{+}z^2-[\omega-{\overline {\overline {\Omega_0}}}{\pm}i(\epsilon+\Gamma)]z
+J_{-}}
\nonumber\\
&&
=-\frac{1}{2\pi i} \oint dz\frac{z^{n-m}}
{J_{+}(z-z_1^{+})(z-z_2^{+})},
\nonumber\\
&&
z_{1}^{+}=
\frac{\omega-{\overline {\overline {\Omega_0}}}{\pm}i(\epsilon+\Gamma)+
\sqrt{[\omega-{\overline {\overline {\Omega_0}}}{\pm}i(\epsilon+\Gamma)]^2-
4J_+J_-}}{2J_{+}},
\nonumber\\
&&
z_{2}^{+}=
\frac{\omega-{\overline {\overline {\Omega_0}}}{\pm}i(\epsilon+\Gamma)-
\sqrt{[\omega-{\overline {\overline {\Omega_0}}}{\pm}i(\epsilon+\Gamma)]^2-
4J_+J_-}}{2J_{+}},
\nonumber\\
&&
z_1^{+}z_2^{+}=\frac{J_{-}}{J_{+}}=\frac
{J-2J^z{\overline {<c_{j{-}1}^+c_j>}}-iD}
{J-2J^z{\overline {<c_{j{+}1}^+c_j>}}+iD}
\end{eqnarray}
for $n\geq m$,
\begin{eqnarray}
&&
{\overline {G_{nm}^{\mp}(\omega{\pm}i\epsilon)}}=-\frac{1}{2\pi i}
\oint dz\frac{z^{\mid n-m \mid}}
{J_{-}z^2-[\omega-{\overline {\overline {\Omega_0}}}{\pm}i(\epsilon+\Gamma)]z
+J_{+}}
\nonumber\\
&&
=-\frac{1}{2\pi i} \oint dz\frac{z^{\mid n-m \mid}}
{J_{-}(z-z_1^{-})(z-z_2^{-})},
\nonumber\\
&&
z_{1}^{-}=
\frac{\omega-{\overline {\overline {\Omega_0}}}{\pm}i(\epsilon+\Gamma)+
\sqrt{[\omega-{\overline {\overline {\Omega_0}}}{\pm}i(\epsilon+\Gamma)]^2-
4J_+J_-}}{2J_{-}},
\nonumber\\
&&
z_{2}^{-}=
\frac{\omega-{\overline {\overline {\Omega_0}}}{\pm}i(\epsilon+\Gamma)-
\sqrt{[\omega-{\overline {\overline {\Omega_0}}}{\pm}i(\epsilon+\Gamma)]^2-
4J_+J_-}}{2J_{-}},
\nonumber\\
&&
z_1^{-}z_2^{-}=\frac{J_{+}}{J_{-}}=\frac
{J-2J^z{\overline {<c_{j{+}1}^+c_j>}}+iD}
{J-2J^z{\overline {<c_{j{-}1}^+c_j>}}-iD}
\end{eqnarray}
for $n \leq m$.
The result of integration can be easily obtained; it depends on the position
of $z_{1,2}^{\pm}$ in complex plane $z$ with respect to unit circle centred
at the origin of complex plane.
The introduced static average fermion correlation functions are given by
\begin{eqnarray}
&&
{\overline {<c_j^+c_j>}}=-\frac{1}{\pi} {\mbox {Im}} \int_{-\infty}^{\infty}
d\omega \frac{{\overline {G_{jj}^{-}(\omega+i\epsilon)}}}
{\exp{(\beta\omega)}+1},\\
&&
{\overline {<c_{j-1}^+c_j>}}={\overline {<c_j^+c_{j+1}>}}=-\frac{1}{\pi}
{\mbox {Im}}
\int_{-\infty}^{\infty}d\omega \frac{{\overline {G_{j,j-1}^{-}
(\omega+i\epsilon)}}}
{\exp{(\beta\omega)}+1},\\
&&
{\overline {<c_{j+1}^+c_j>}}={\overline {<c_j^+c_{j-1}>}}=-\frac{1}{\pi}
{\mbox {Im}}
\int_{-\infty}^{\infty}d\omega \frac{{\overline {G_{j,j+1}^{-}
(\omega+i\epsilon)}}}
{\exp{(\beta\omega)}+1}.
\end{eqnarray}

The equations (4.8)-(4.12) permit to determine self-consistently
correlation functions
${\overline {<c_j^+c_j>}},\:{\overline {<c_{j+1}^+c_j>}},\:{\overline
{<c_{j-1}^+c_j>}}$
that were introduced in (4.4). Knowing them one can get the average
spectral density that is given by
\begin{eqnarray}
{\overline {\rho (E)}}=
\nonumber\\
-\frac{1}{\pi}{\mbox {Im}} \frac{1}{2\pi}\int_{-\pi}^{\pi}
d\kappa
\frac{1}{\omega{-}\left [{\overline {\overline {\Omega_0}}}
{+}J_+\exp{(i\kappa)}
{+}J_-\exp{(-i\kappa)}\right ]{+}i(\epsilon{+}\Gamma)}
\nonumber\\
=\frac{1}{\pi}{\mbox {Im}} \frac{1}{2\pi i}\oint \frac{dz}
{J_{\pm}(z-z_1^{\pm})(z-z_2^{\pm})}.
\end{eqnarray}
Such approach can be viewed as a
generalization of rather old ideas (see \cite{b}) for the case of
random lorentzian external field.

\section{
Conclusions}

The results which have been obtained in the present paper are summarized as
follows. Considering spin-$\frac{1}{2}$ isotropic $XY$ chain with
Dzyaloshin\-skii-Moriya interaction in random lorentzian transverse field
(1.1) as a system of fermions (1.3) one is able to obtain exactly the average
one-fermion Green's functions
${\overline {G_{nm}^{\pm}(\omega{\pm}i\epsilon)}}$
(1.21). This permits to get exact result for average spectral density (1.22)
and therefore, for thermodynamical quantities (2.2)-(2.5), (2.7)-(2.10), to
examine the influence of Dzyaloshinskii-Moriya interaction on them
(Figs.1-5). Unfortunately, it appeared impossible to reach any exact results
concerning spin correlations in model in question. The system that is
considered does not represent accurately any physical system, but may serve
as a testing ground for theory and numerical techniques and in this sence,
has proved most valuable. The results of comparison of conclusions obtained
within different approximate approaches with exact ones (Figs.6,7,
Fig.8) represent some interest and may be useful in
understanding the region of validity of various
well-known methods of randomly disordered crystals theory.
After introducing into the Hamiltonian of the interaction between $z$
components of neighbouring spins one is unable to obtain exact results since
these terms lead to appearance of products of four Fermi operators in (4.2).
After adopting approximation (4.4) the problem reduces to already
considered one, however, more complicated because of necessity to determine
self-consistently three fermion correlation functions.

The performed exact calculations for the system with Hamiltonian (1.1) or
(1.3) are anticipated to be useful in leading to an
understanding of the properties of one-dimensional disordered models of
condensed matter physics.

\section*{
Acknowledgements}

The authors appreciate the advices and interest in this work of many
colleagues,
but particular thanks go to Dr. T.Krokhmalskii, Dr. V.Tkachuk,
and Mr. Yu.Dublenytch.
The authors also would like to express their gratitude to the participants of
the seminar of Laboratory for the Theory of Model Spin Systems of Quantum
Statistics Department of ICMP (20.05.1994), of the seminar
of Quantum Statistics Department of ICMP (24.05.1994) and of the seminar at
the
Chair of Theoretical Physics of Ivan Franko Lviv State University
(15.06.1994)
for many helpful discussions.

\end{document}